\documentclass[9pt,conference]{IEEEtran}
\IEEEoverridecommandlockouts
\usepackage{cite}
\usepackage{amsmath,amssymb,amsfonts}
\usepackage{algorithmic}
\usepackage{graphicx}
\usepackage{textcomp}
\usepackage{balance}
\usepackage{xcolor}
\usepackage{comment}
\usepackage{hyperref}

\def\BibTeX{{\rm B\kern-.05em{\sc i\kern-.025em b}\kern-.08em
    T\kern-.1667em\lower.7ex\hbox{E}\kern-.125emX}}
\begin{document}

\newcommand{\expnumber}[2]{{#1}\mathrm{e}{#2}}

\title{MambaFoley: Foley Sound Generation using Selective State-Space Models}

\author{\IEEEauthorblockN{Marco Furio Colombo, Francesca Ronchini, Luca Comanducci, Fabio Antonacci} \IEEEauthorblockA{
Dipartimento di Elettronica, Informazione e Bioingegneria (DEIB), Politecnico di Milano\\
Piazza Leonardo Da Vinci 32, 20133 Milan, Italy\\
Email: marcofurio.colombo@mail.polimi.it, francesca.ronchini@polimi.it, luca.comanducci@polimi.it, fabio.antonacci@polimi.it
}
{\thanks{This work was partially supported by the European Union under the Italian National Recovery and Resilience Plan (NRRP) of NextGenerationEU (PE00000014 - program ``SERICS'').}}

}

\maketitle

\begin{abstract}
Recent advancements in deep learning have led to widespread use of techniques for audio content generation, notably employing Denoising Diffusion Probabilistic Models (DDPM) across various tasks.  Among these, Foley Sound Synthesis is of particular interest for its role in applications for the creation of multimedia content. 
Given the temporal-dependent nature of sound, it is crucial to design generative models that can effectively handle the sequential modeling of audio samples. Selective State-Space Models (SSMs) have recently been proposed as a valid alternative to previously proposed techniques, demonstrating competitive performance with lower computational complexity.
In this paper, we introduce MambaFoley, a diffusion-based model that, to the best of our knowledge, is the first to leverage the recently proposed SSM known as Mamba for the Foley sound generation task.
To evaluate the effectiveness of the proposed method, we compare it with a state-of-the-art Foley sound generative model using both objective and subjective analyses. 
\end{abstract}

\begin{IEEEkeywords}
Foley Sound Synthesis, State-Space Models, Generative Sound Synthesis, Denoising Diffusion Probabilistic Models
\end{IEEEkeywords}

\section{Introduction}
In recent years, audio generation using deep-learning generative techniques has made relevant progress. This was possible primarily by the adoption of generative models denoted as Denoising Diffusion Probabilistic Models (DDPMs)~\cite{ho2020denoising}, which have been applied to a wide variety of audio applications such as Text-to-Audio generation~\cite{yang2023diffsound,liu2024audioldm,huang2023make}, 
sound field reconstruction~\cite{miotello2024reconstruction}, among others. 
Generated audio samples from generative models are so realistic that using them to replace real datasets for training deep learning models has been explored across various domains~\cite{feng2024can,ronchini2024synthetictrainingsetgeneration}. 

These advancements in audio generation also extend to Foley sound synthesis, which refers to the task of creating realistic audio excerpts for enhancing the quality of multimedia content during post-production, (e.g. gunshot sounds in movies).
Traditionally, this task was performed by human Foley artists. In 2023, the introduction of a dedicated task in the DCASE challenge~\cite{choi2023foley} encouraged the adoption of generative models for automatic Foley sound creation. This led to the development of models primarily based on diffusion-based techniques~\cite{kang2023fall,scheiblerfoley}, including methods for generating sounds controlled by textual input~\cite{yuan2023text}, as well as solutions using Generative Adversarial Networks (GANs)~\cite{chungfoley}. Although the quality of the generated sounds is impressive, the proposed models face several challenges, such as the precise control of the timing of synthesized sound onsets. Over the past year, various Foley generative models have started addressing this challenge. SyncFusion~\cite{comunita2024syncfusion} generates synchronized sound effect tracks by extracting event onsets from videos and using them to condition diffusion models. T-Foley~\cite{chung2024t}, on the other hand, provides conditioning through the Block-FiLM technique, which allows the diffusion model to be controlled via a temporal feature. More recently, PicoAudio~\cite{xie2024picoaudio} introduced a text-conditioned diffusion model for Foley sound generation, achieving notable results through the development of a curated dataset with accurately annotated Foley sound event timestamps. The same authors proposed an evaluation dataset specifically designed for time-controlled sound event generation~\cite{xie2024audiotime}.

\begin{figure}[t!]
   \centering  \centerline{\includegraphics[width=\columnwidth]{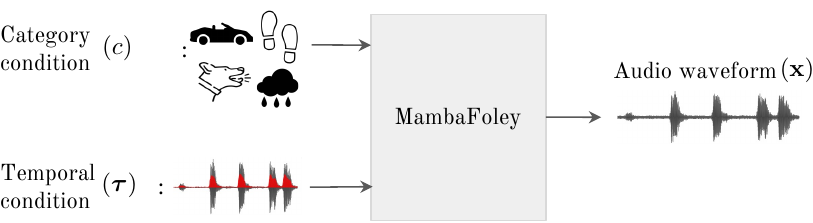}}
   \caption{Schematic representation of MambaFoley inference procedure. The model generates a raw audio waveform based on the desired input category and temporal profile.
   }
   \label{fig:model_scheme}
\end{figure}

Generating audio waveforms with mid and long-term temporal coherence has always been a challenge for learning-based models. In fact, although using raw waveforms in audio generative modeling can significantly improve the quality of generated audio content, they also pose challenges due to the high-dimensional input signal, which includes tens of thousands of samples. Several deep learning models have been proposed to address the challenge of modeling raw waveform sequences, each with its own drawbacks~\cite{mehri2022samplernn, van2016wavenet, child2019generating, vaswani2017attention}. RNNs based on LSTMs~\cite{mehri2022samplernn} are not parallelizable, which limits their computational efficiency. CNNs~\cite{van2016wavenet} are constrained by their receptive field, which limits their ability to model the global coherence of the waveform. Transformers and, in general, attention-based methods~\cite{child2019generating, vaswani2017attention}, while effective for modeling sequential data, are impractical for very long sequences due to their quadratic complexity with respect to window length. Recently, structured State-Space Models (SSMs)~\cite{gu2021combining,guefficiently} have been proposed as a new class of architectures tailored to sequence modeling, combining characteristics of both CNNs and RNNs. Architectures such as SaShiMi~\cite{goel2022s} and especially Mamba~\cite{mamba,mamba2} outperformed previous state-of-the-art techniques over waveform modeling while scaling linearly in complexity with respect to sequence length. As expected, Mamba models were readily applied to a diverse variety of audio signal processing tasks such as audio representation learning~\cite{yadav2024audio}, deepfake detection~\cite{chen2024rawbmamba}, and speech separation~\cite{SPMamba}. 
Motivated by previous studies, we explored how to integrate selective SSMs into the Foley sound generation task.

In this paper, we introduce MambaFoley, a method for Foley sound generation that leverages the SSM Mamba to model raw audio. To the best of our knowledge,
this is the first application of Mamba in Foley sound generation. 
The proposed model employs the diffusion architecture DAG~\cite{DAG} and integrates Mamba in the bottleneck to improve its sequence modeling capabilities. Additionally, we enable temporal control of the generated Foley sounds through the BlockFiLM conditioning method introduced in~\cite{chung2024t}. 
Fig.~\ref{fig:model_scheme} shows a schematic overview of the MambaFoley model. To validate the effectiveness of our proposed model, we conduct a series of experiments comparing it with the model proposed in~\cite{chung2024t} and an Attention-based baseline. We evaluate performance using both objective metrics and subjective evaluations. The rest of the paper is organized as follows: Sec.~\ref{sec:background} provides the reader with the necessary background on SSMs. In Sec.~\ref{sec:method}, we present design choices of the MambaFoley model. Sec.~\ref{sec:expsetup} describes how the experiments were organized, while Sec.~\ref{sec:results} presents the results, highlighting the effectiveness of the proposed technique in comparison to state-of-the-art methods. In Sec.~\ref{sec:conclusion} we draw some conclusions. 
Code and listening examples are publicly available online\footnote{\href{https://furiocolombo.github.io/mamba-foley}{https://furiocolombo.github.io/mamba-foley}}.

\section{State-Space Models background}
\label{sec:background}
This section provides a brief overview of SSMs; for more comprehensive details, we refer the reader to~\cite{gu2021combining,guefficiently,goel2022s,mamba,mamba2}. 

SSMs are traditionally defined as mappings in the continuous domain $u(t)\in \mathbb{R} \mapsto y(t) \in \mathbb{R}$, with a hidden state $h(t) \in \mathbb{R}^D$, characterized by system:
\begin{equation}
h'(t) = \mathbf{A}h(t) + \mathbf{B}u(t);\quad
y(t) = \mathbf{C}h(t) + \mathbf{D}u(t),
\end{equation}
where $\mathbf{A} \in \mathbb{R}^{D\times D}$ is the state-transition matrix, $\mathbf{B}\in \mathbb{R}^{D\times 1}$ the input projection matrix, $\mathbf{C} \in \mathbb{R}^{1\times D}$ the output projection matrix and $\mathbf{D}$ the feedthrough matrix. The feedthrough matrix $\mathbf{D}$ will be omitted in the rest of the paper, as it is straightforward to compute due to its direct dependence on the input~\cite{guefficiently}.

To adapt to discrete input sequences, the system needs to be discretized by a step size $\Delta$, which corresponds to the input resolution. The system matrices can be recomputed using zero-order hold (ZOH) discretization~\cite{mamba} as follows:
\begin{equation}
    \overline{\mathbf{A}} = \exp{(\Delta\mathbf{A})}; \quad \overline{\mathbf{B}} = (\Delta \mathbf{A})^{-1} (\exp{(\Delta\mathbf{A})-\mathbf{I}})\Delta\mathbf{B},
\end{equation}
resulting in the discretized system:
\begin{equation}
\label{eq:discrete_system_SSM}
    h_t = \overline{\mathbf{A}}h_{t-1} + \overline{\mathbf{B}}u_t;\quad
y_t = \mathbf{C}h_t.
\end{equation}

The State-Space model in \eqref{eq:discrete_system_SSM} can then be computed as a global convolution between input sequence $\mathbf{u} =[u_0, u_1, \ldots, u_{T-1}]$ and a kernel $\overline{\mathbf{K}} \in \mathbb{R}^M$: 
\begin{equation}
    \overline{\mathbf{K}} = (\mathbf{C}\overline{\mathbf{B}}, \mathbf{C}\overline{\mathbf{A}\mathbf{B}},\ldots,\mathbf{C}\overline{\mathbf{A}}^{T-1}\overline{\mathbf{B}}),
\end{equation}
where $T$ is the length of the input sequence $\mathbf{u}$ in samples and finally
\begin{equation}
    y_t = \mathbf{u} * \overline{\mathbf{K}},
\end{equation}
where $*$ denotes convolution.
The kernel $\overline{\mathbf{K}}$ can be pre-computed when all parameters are fixed, as in models like SaShiMi~\cite{goel2022s}. However, in \textit{selective} SSMs like Mamba, the parameters $\Delta_t$, $\overline{\mathbf{A}}_t$, $\overline{\mathbf{B}}_t$ and $\mathbf{C}_t$ are time-dependent and update at each timestep $t$ 
based on the input $x_t$.

\section{MambaFoley}
\label{sec:method}
This section introduces the proposed generative model,  MambaFoley. First, we present the Foley sound synthesis generation task, followed by a description of the overall model framework, including the details of the network architecture.

\subsection{Problem formulation}
The goal of MambaFoley is to generate raw audio waveforms that represent specific sound classes (e.g. dog, gunshot, etc.) while also following a specified temporal pattern (e.g. two dog barks separated by one second of silence).

More formally, let $\mathbf{x} \in \mathbb{R}^{N}$ be a discrete audio waveform whose content corresponds to the category indicated by the class $c \in \{0,\ldots, C-1\}$ of length $N$ samples We want to control the temporal evolution of $\mathbf{x}$ using $\boldsymbol{\tau} \in \mathbb{R}^{I}$, which can be defined using different representations.


MambaFoley then learns the function $\mathcal{M}(\cdot)$ such that $ \mathbf{x} = \mathcal{M}(c,\boldsymbol{\tau}).$
Following the approach proposed in~\cite{chung2024t}, we condition the temporal evolution of the generated audio using the Root-Mean-Square (RMS) of a waveform signal that contains an audio excerpt of the target category with the desired temporal evolution.
Specifically, in our scenario we can define:  
\begin{equation}
\tau[i] = \sqrt{\frac{1}{W} \sum_{n=iH}^{iH+W} x^2[n]},
\end{equation}
where $i$, $W$ and $H$ represent the frame index, window size (in samples) and hop size (in samples), respectively.

\begin{figure}[t!]
   \centering  \centerline{\includegraphics[width=\columnwidth]{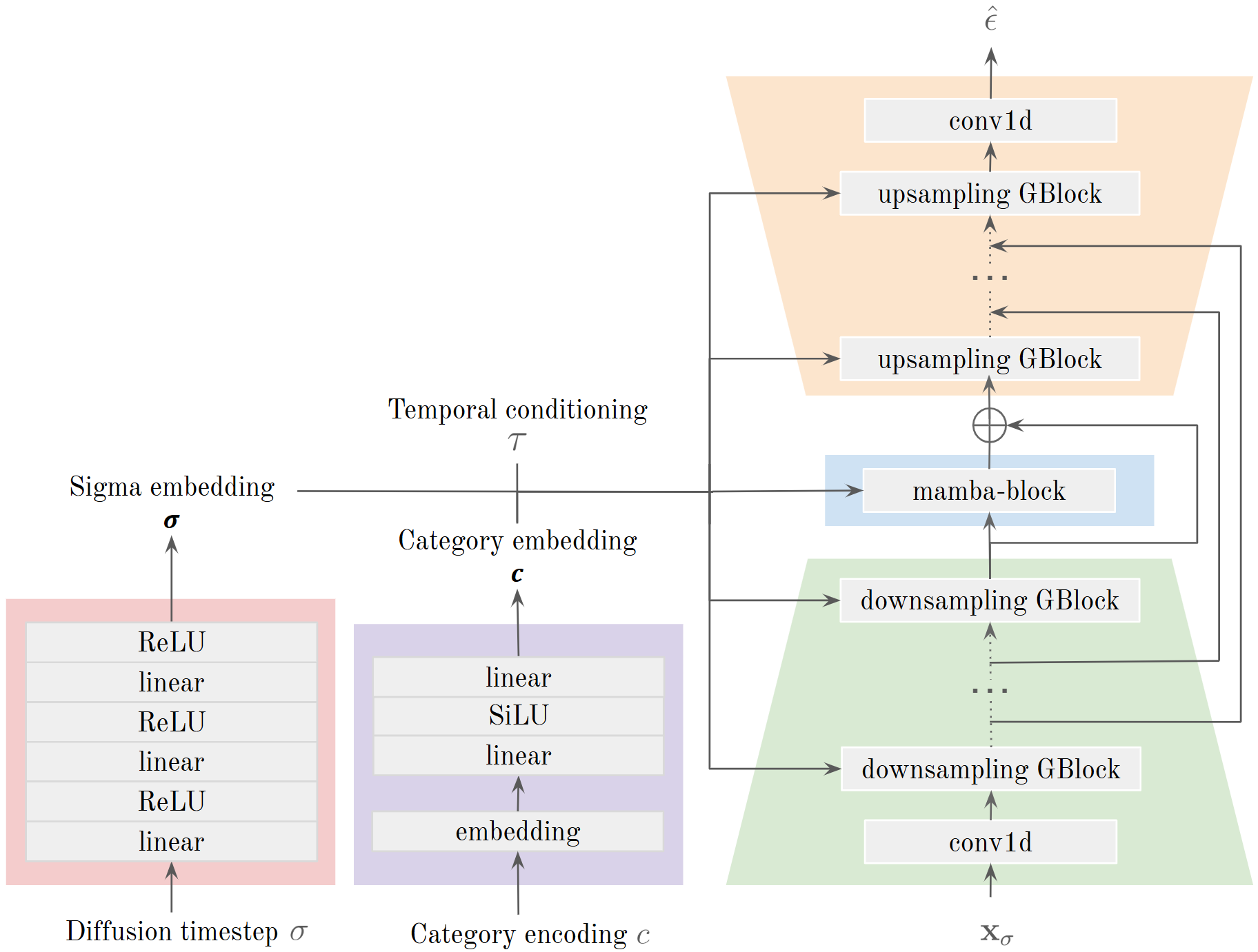}}
    \caption{Schematic representation of the U-Net architecture used during the backward part of the diffusion process.}
    
   \label{fig:unet-architecture}
\end{figure}

\subsection{Network Architecture}
MambaFoley is a diffusion-based model~\cite{ho2020denoising}, where the target audio waveform $\mathbf{x}$ is progressively corrupted with Gaussian noise throughout the forward process. In the backward process, the original waveform is iteratively denoised using a U-Net architecture, whose parameters are updated based on the $\ell_2$ norm between the predicted and ground truth noise $\boldsymbol{\epsilon}$. The overall architecture of MambaFoley is illustrated in Fig.~\ref{fig:unet-architecture}.

We select the DAG architecture~\cite{DAG} as the U-Net backbone of our model, due to its proven effectiveness in generating high-resolution raw audio. We introduce some changes to DAG, specifically regarding the conditioning and bottleneck of the model. 

\begin{figure}[t]
\centering
\begin{minipage}[c]{0.99\columnwidth}
  \centering
\centerline{\includegraphics[width=\columnwidth]{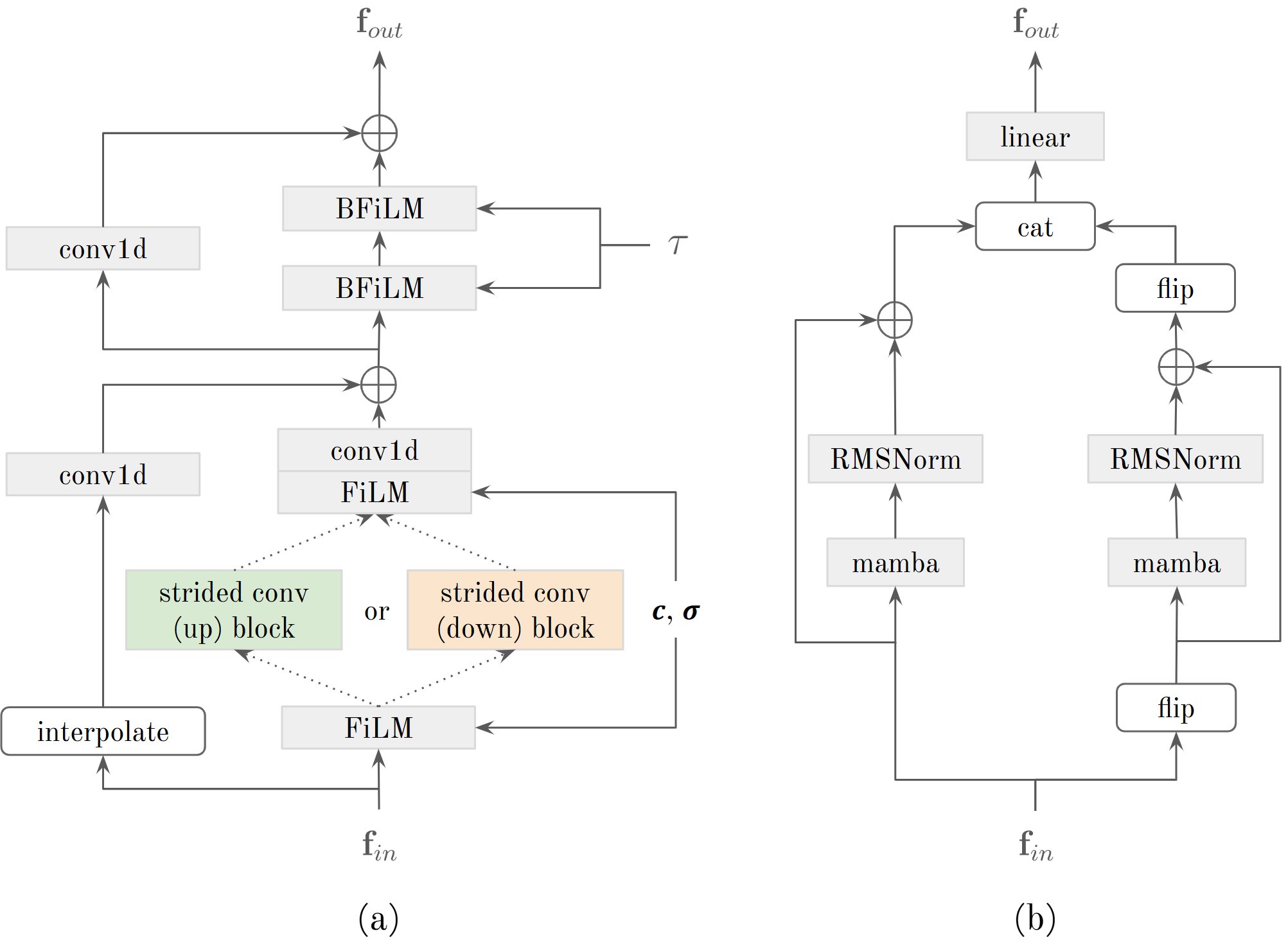}}
\end{minipage}

\caption{Layers of MambaFoley: GBlock (a) and bidirectional Mamba bottleneck (b), $\mathbf{f}_{in}$ and $\mathbf{f}_{out}$ represent the generic input and output feature of the layers, respectively.}
\label{fig:mamba_foley_layers}
\end{figure}

Similar to DAG, we refer to the building blocks of MambaFoley as \textit{GBlocks}, to which we apply significant modifications. GBlocks consist of sequences of 1D-convolutional layers in the downsampling encoder part, or transposed 1D-convolutions in the upsampling decoder. Fig.~\ref{fig:mamba_foley_layers} (a) shows the architecture of GBlocks. 

The conditioning of the proposed model is performed in the GBlocks, by injecting the categorical information via FiLM~\cite{Perez_Strub_deVries_Dumoulin_Courville_2018FiLM} layers and the temporal information via BFiLMs~\cite{chung2024t}.

In the bottleneck of the model, between the encoder and decoder parts of the U-Net, we introduce a bidirectional Mamba layer~\cite{SPMamba, hwang2024hydra} to exploit the capabilities of SSMs in processing temporal audio sequences. This layer consists of two parallel Mamba blocks, as defined in Sec.~\ref{sec:background}, one processing the input sequence and the other one processing its reversed version.  Fig.~\ref{fig:mamba_foley_layers} (b) shows the bidirectional Mamba bottleneck. The outputs of the two parallel Mamba blocks are each normalized using Root Mean Square Layer Normalization (RMSNorm)~\cite{zhang2019root}, chosen for its computational efficiency. The normalized outputs are then summed with the input (and reversed) signal residuals. Finally, the outputs of the two SSMs are concatenated and processed by a single linear layer.
We use state dimension D=$16$, expansion factor $4$ and local convolution width $4$ for the Mamba layers.


\section{Experimental Setup}
\label{sec:expsetup}
This section describes the evaluation setup used for the experiments. First, we present the dataset considered for the evaluation, followed by the baseline models selected for comparison with MambaFoley and details of the training setup. Finally, we explain the experimental procedures for both objective and subjective evaluations.

\subsection{Dataset}
We consider the same dataset provided for the DCASE 2023 Task 7, corresponding to the Foley Sound Synthesis Challenge~\cite{choi2023foley}. Specifically, the dataset contains a total of $6.1$ hours of audio annotated across seven distinct categories, selected to ensure a balanced representation across \textit{human}, \textit{nature}, and \textit{mechanical} sound classes. 
The dataset is split into development and evaluation sets, with the evaluation set containing 100 audio samples per category, corresponding to approximately $14\%$ of the total dataset.
The audio clips are sourced from UrbanSound8K~\cite{SalamonUrbanSound14}, FSD50K~\cite{fonseca2021fsd50k}, and BBC Sound Effects~\cite{BBCSoundEffects}. All the audio clips are standardized to mono $16$-bit $4$-second snippets at a sampling rate of $22,050$ Hz.

\subsection{Techniques Under Comparison}
We consider two distinct baselines as comparison models for the proposed techniques. 

We compare MambaFoley with T-Foley~\cite{chung2024t}, one of the most relevant methods in the literature. T-Foley is a temporal-guided approach for Foley sound synthesis that also employs DAG~\cite{DAG} as its audio generation backbone along with BFiLM conditioning. Differently from our model, T-Foley uses LSTMs in the bottleneck of the U-Net, as in the original DAG paper~\cite{DAG}. 

To prove the effectiveness of the adoption of SSM audio waveform generation against other sequence modeling methods, we create a custom baseline considering attention-based layers. Specifically, we modify MambaFoley by replacing the bidirectional Mamba layer with a multihead attention layer with $4$ parallel heads, each with a dimension of $64$. 
For the rest of the paper, we will further refer to this architecture as \textit{AttentionFoley}.

\subsection{Training details}
\label{subsec:expdetails}
 All models are trained for $500$ epochs using the Adam~\cite{diederik2014adam} optimizer with a learning rate of 1e-4. To make the training more robust, we randomly train the network unconditionally with a $0.1$ probability. We adopt a variance-preserving cosine scheduler, as proposed in ~\cite{rouard2021crashrawaudioscorebased}. For conditional sampling, we employ the DDPM-like discretization of SDE~\cite{VariationalDiffModelsDDPM} with classifier-free guidance~\cite{ho2022classifierfreediffusionguidance} to enhance sample diversity and quality.
For the temporal event features, we use a window length of $W=512$ samples and a hop size of $H=128$ samples, which correspond to approximately $24~\mathrm{ms}$ and $6~\mathrm{ms}$, respectively, given the sampling rate of $22.05$ kHz. For a fair comparison with T-Foley~\cite{chung2024t}, we also exclude the \textit{SneezeCough} class from the listening test.

\begin{figure*}[htb]
\centering
\begin{minipage}[c]{.2\textwidth}
  \centering
\centerline{\includegraphics[width=\textwidth]{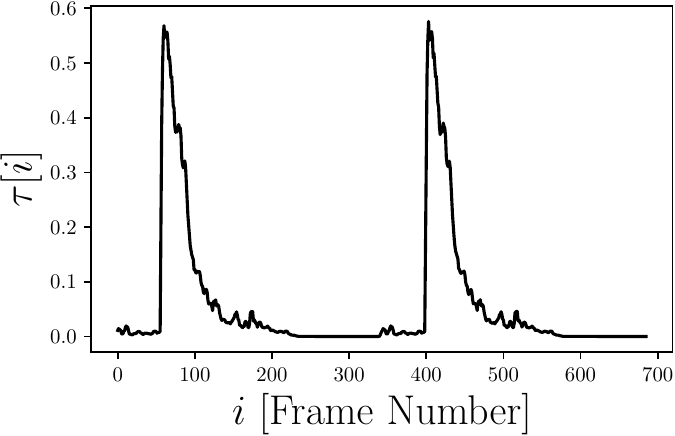}}
\centerline{(a) Temporal Conditioning}
\end{minipage}
\hfill
\begin{minipage}[c]{0.19\textwidth}
  \centering
\centerline{\includegraphics[width=\textwidth]{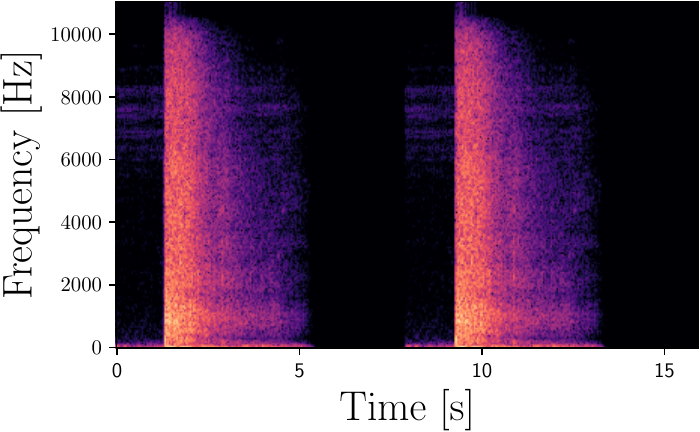}}
\centerline{(b) Target output}
\end{minipage}
\hfill
\begin{minipage}[c]{0.19\textwidth}
  \centering
\centerline{\includegraphics[width=\textwidth]{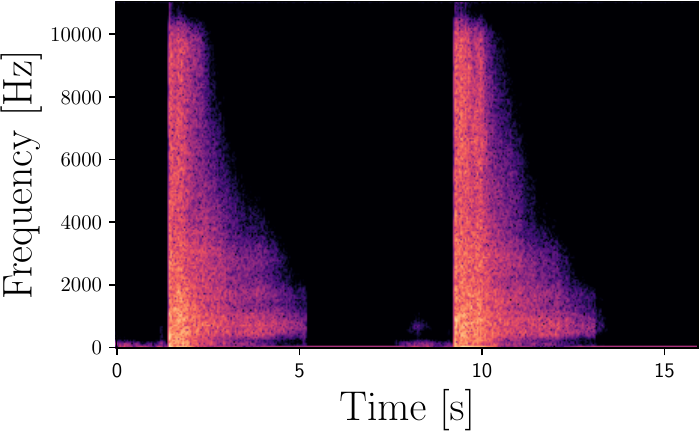}}
\centerline{(c) MambaFoley}
\end{minipage}
\hfill
\begin{minipage}[c]{0.19\textwidth}
  \centering
\centerline{\includegraphics[width=\textwidth]{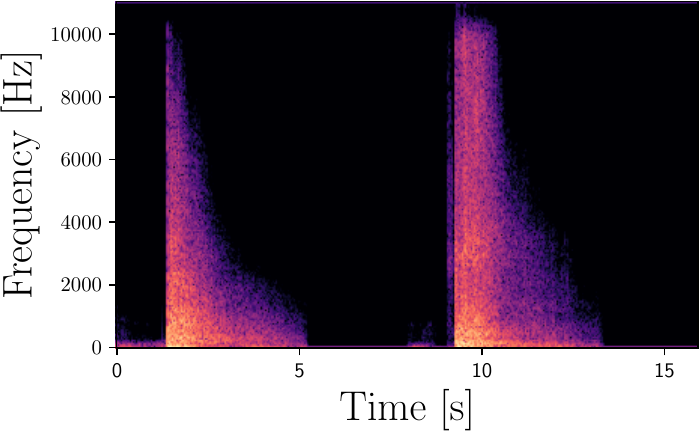}}
\centerline{(d) T-Foley}
\end{minipage}
\hfill
\begin{minipage}[c]{0.19\textwidth}
  \centering
\centerline{\includegraphics[width=\textwidth]{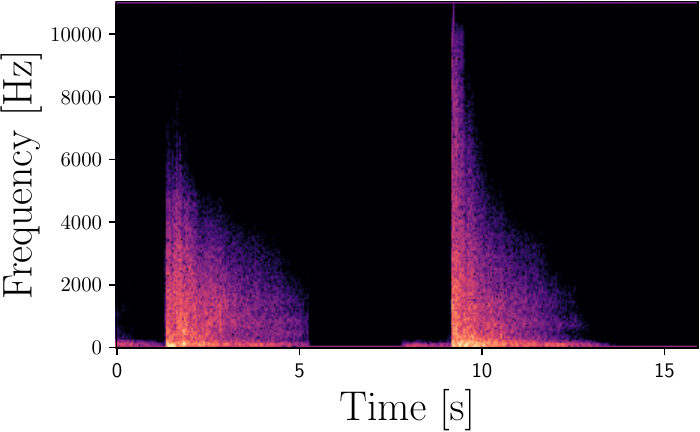}}
\centerline{(e) AttentionFoley}
\end{minipage}
\caption{
{
Example spectrograms (c-e) of samples generated using different models.} The category conditioning corresponds to the class \textit{gunshot}, while the temporal conditioning is provided via the RMS shown in (a) and computed over the ground truth shown in (b).
}
\label{fig:spec_examples}
\end{figure*}

\subsection{Objective evaluation}
\label{subsec:objective_eval}
We consider two objective metrics to evaluate the generated audio samples, the Fréchet Audio Distance (FAD)~\cite{kilgour2019frechetaudiodistancemetric} and E-L1 distance. 

FAD has been proposed in~\cite{kilgour2019frechetaudiodistancemetric} for music evaluation and it was adapted from the Frechet Inception Distance (FID) used in the image domain~\cite{NIPS2017_FID}. The FAD metric evaluates the distributional similarity between two datasets within a given embedding space, with the VGGish model~\cite{hershey2017cnn} initially proposed as the feature extractor. FAD has been widely adopted for evaluating generative models, but recent studies have shown that the FAD correlation with human perception is embedding-dependent~\cite{tailleur2024correlation}  and that VGGish may not always be the most effective feature extractor~\cite{tailleur2024correlation, gui2024adapting}. To address these concerns, and following the approach proposed in~\cite{chung2024t}, we use multiple feature extractors. Specifically, we compute the FAD using VGGish~\cite{hershey2017cnn} and two PANN models~\cite{panns}: PANNs-CNN14-16k and PANNs-CNN14-32k. 
We will refer to FAD values computed using the VGGish model as FAD-V, while those calculated with PANNs-CNN14-16k and PANNs-CNN14-32k as FAD-P16 and FAD-P32, respectively. To compute the FAD values, we utilized the toolkit\footnote{\href{https://github.com/DCASE2024-Task7-Sound-Scene-Synthesis/fadtk}{https://github.com/DCASE2024-Task7-Sound-Scene-Synthesis/fadtk}} developed in~\cite{gui2024adapting}.

E-L1 distance is a metric to assess the temporal alignment of the generated audio and the temporal conditioning. We adopted the same implementation described in~\cite{chung2024t}.
Formally, given the desired temporal conditioning $\tau$ and the one estimed from the generated waveform $\hat{\tau}$, the E-L1 distance is the E-L1 distance is defined as:
\begin{equation}
 E_{L1}(\tau, \hat{\tau}) = \frac{1}{T} \sum_{i=1}^{N} |\tau_i - \hat{\tau}_i|
\end{equation}
 where $x_i$ and $\hat{x}_i$ represent the $i$-th sample of the original and generated audio signals, respectively.
 
\subsection{Subjective evaluation}
\label{subsec:subjective_eval}
We conduct a subjective listening evaluation test to assess the Overall Quality and Temporal Fidelity of the audio generated by the proposed model. The Overall Quality assesses the perceptual quality of the generated samples, whereas the Temporal Fidelity section evaluates how closely the temporal features of the generated audio align with the conditioning. Participants for the listening test were recruited through mailing lists from the broader audio research community. The test duration was approximately 10 minutes. No personal data was collected. A total of 20 participants were asked to rate the proposed audio samples on a 5-point Likert scale, where 1 represents poor quality or low temporal alignment and 5 signifies excellent quality or near-perfect temporal adherence (according to the test section).
In this paper, the results are reported using the Mean Opinion Score (MOS), based on evaluations gathered from test participants.

\section{Results}
\label{sec:results}
In this section, we present results demonstrating the effectiveness of the proposed MambaFoley model compared to T-Foley and AttentionFoley baselines, both in terms of objective and subjective evaluation metrics. Before presenting the analyses computed on the entire dataset, we examine a representative spectrogram generated from the waveforms obtained through each of the considered models, shown in Fig.~\ref{fig:spec_examples}. The STFT for all spectrograms is computed with a Hann window of length $512$, hop size of $128$ samples and a FFT of $512$ points. Fig.~\ref{fig:spec_examples}(b) contains an example of the target sound, corresponding to a \textit{gun shot}, from which we extract the RMS, shown in Fig.\ref{fig:spec_examples} (a). The RMS is then used to condition the temporal evolution of the network. As observed from the spectrograms of all the considered models in Fig.~\ref{fig:spec_examples}(c) (MambaFoley), (d) (T-Foley), and (e) (AttentionFoley), each model appears to accurately reproduce the desired temporal conditioning. However, MambaFoley seems to reconstruct more content at higher frequencies.
\subsection{Objective Evaluation Results}

 

\begin{table}[t!]
\centering
    \caption{Objective Evaluation Results in Terms of E-L1 and FAD.}
    \resizebox{\columnwidth}{!}{
    \begin{tabular}{lcccccc}
        \hline
        Model & \#par$\downarrow$ & FLOPs$\downarrow$ & Inf. Time$\downarrow$ & E-L1$\downarrow$ & FAD-P32$\downarrow$ & FAD-V$\downarrow$ \\
        \hline
        Real data       & -   & -                      & -              & 0                 & 15.20             & 3.31          \\
        \hline
        T-Foley         & 74M & $\expnumber{5.42}{10}$ & 25.95          & \textbf{0.0351}   & 40.41             & 8.60          \\
        AttentionFoley  & 63M & $\expnumber{5.38}{10}$ & 16.84          & 0.0393            & 40.58             & 8.35          \\
        MambaFoley      & 70M & $\expnumber{5.40}{10}$ & \textbf{16.48} & 0.0374            & \textbf{35.43}    & \textbf{7.77} \\
        \hline
        \end{tabular}}
\label{tab:obj-eval-comparison}
\end{table}

Table~\ref{tab:obj-eval-comparison} reports the results for E-L1 and FAD for the three different feature extractors considered in this study. It also includes  the number of parameters for each network, required floating-point operations per second (FLOPs) and the generation time for a single audio sample.
To facilitate the analysis, we highlight the best result in each category in bold. The top row presents the lower bound on the achievable error, represented by the FAD and E-L1 scores computed on real data.
Although the lower bound of the E-L1 distance is clearly 0, the FAD calculated between the embeddings of the test and train datasets for real data is greater than zero. This indicates the inherent differences between real audio samples. This finding helps to estimate the lower bound for FAD in our task, where our goal is to generate as realistic audio samples as possible.

Table~\ref{tab:obj-eval-comparison} shows that MambaFoley achieves faster inference and outperforms both T-Foley and AttentionFoley in terms of FAD. The models demonstrate comparable E-L1 performance, likely due to the influence of the conditioning blocks over the bottleneck, which help maintain consistent temporal alignment.

\subsection{Subjective Evaluation Results}
\begin{table}[t!]
\centering
    \caption{
    Subjective Evaluation in Terms of MOS. We report the average and 95\% confidence intervals.
    }
    \vspace{-0.3em}
    \resizebox{\columnwidth}{!}{\begin{tabular}{lccccc}
        \hline
        Model               & \hspace{0.1cm} & Overall Quality (MOS)$\uparrow$ & \hspace{0.2cm} & Temporal Fidelity (MOS)$\uparrow$ \\
        \hline
         T-Foley            & \hspace{0.1cm} & $3.18\ (\pm 0.41)$           & \hspace{0.2cm} & $3.45\ (\pm 0.35)$           \\
         AttentionFoley     & \hspace{0.1cm} & $2.76\ (\pm 0.65)$           & \hspace{0.2cm} & $\textbf{3.68}\ (\pm 0.39)$  \\
         MambaFoley         & \hspace{0.1cm} & $\textbf{3.75}\ (\pm 0.53)$  & \hspace{0.2cm} & $\textbf{3.68}\ (\pm 0.41)$           \\
        \hline
        \end{tabular}
        }
\label{tab:subj-eval-comparison}
\end{table}

Table~\ref{tab:subj-eval-comparison} presents the subjective evaluation results averaged over the 20 participants. Results indicate that MambaFoley is perceived as superior to the other methods in terms of \textit{Overall Quality}, a measure that encompasses various aspects primarily related to the timbral perception of the sound. This confirms the hypotheses drawn out while looking at the spectrograms in Fig.~\ref{fig:spec_examples}. Regarding \textit{Temporal Fidelity}, the models appear more similar in quality, although perceptual results slightly favor MambaFoley and AttentionFoley. This may be attributed to the use of BFiLM temporal conditioning blocks, which are shared among all the compared models. Overall, the results suggest that different sequence modeling techniques are more crucial for capturing the correct timbral aspects of the sounds rather than modeling the temporal evolution, while maintaining other components of the network similar.

\section{Conclusion}
\label{sec:conclusion}
In this paper, we introduced MambaFoley, a Denoising Diffusion Probabilistic model for Foley sound generation exploiting the power of selective State-Space models. Specifically, we have applied Mamba in the bottleneck of a diffusion model, a SSM that has proved to obtain notable performance in audio sequence modeling, while mantaining computational complexity. Through a series of experiments that include both objective and subjective results, we demonstrate the effectiveness of the proposed approach by comparing it with baseline methods using various sequence modeling techniques. Results motivate us to explore and optimize the proposed approach further to enhance its performance in generative audio tasks.

\ifCLASSOPTIONcaptionsoff
  \newpage
\fi
\balance
\bibliographystyle{IEEEtran}
\bibliography{biblio}

\end{document}